# FAULT TOLERANT CONSENSUS AGREEMENT ALGORITHM

Marius Rafailescu[1]

[1]The Faculty of Automatic Control and Computers, POLITEHNICA University, Bucharest

marius.rafailescu@cti.pub.ro

## ABSTRACT

*Recently a new fault tolerant and simple mechanism was designed for solving commit consensus problem. It is based on replicated validation of messages sent between transaction participants and a special dispatcher validator manager node. This paper presents a correctness, safety proofs and performance analysis of this algorithm.*

## KEYWORDS

*consensus agreement, fault tolerance, leader election, distributed systems*

## 1. INTRODUCTION

Consensus algorithms were discussed in the past and several solutions where developed (Two-phase commit, Three-phase commit or Paxos) [1]. The latter is fault tolerant and with the introduction of distributed databases it was implemented in many systems, although it is not so easy to implement [2]. In recent years was defined a new algorithm, Raft, which has been developed in order to provide a consensus control for replicated state machines, intended to be easy to understand and implement [3].

One recent work [4] describes a new algorithm which uses a set of validator nodes, including one dispatcher and also presents algorithm for dispatcher election (equivalent to leader election) made for the purpose of not rollbacking the transaction when a new dispatcher is chosen. Based on this new consensus agreement solution, this paper highlights the correctness and safety proofs of the algorithm.

## 2. DESCRIPTION

The system is modelled in an asynchronous way (with the corresponding implication of using timeouts - as a well known result [5]), having the following suppositions:

- Messages can take an arbitrary number of steps from source to destination;
- Messages can be reordered, duplicated or lost by the network, but never corrupted;
- Nodes fail by stopping; later, they can restart and re-enter in the system.

The specification describes a system with an arbitrary number of nodes, which communicate through messages which are sent in two manners: one-to-one and one-to-many as we can see in figure 1.

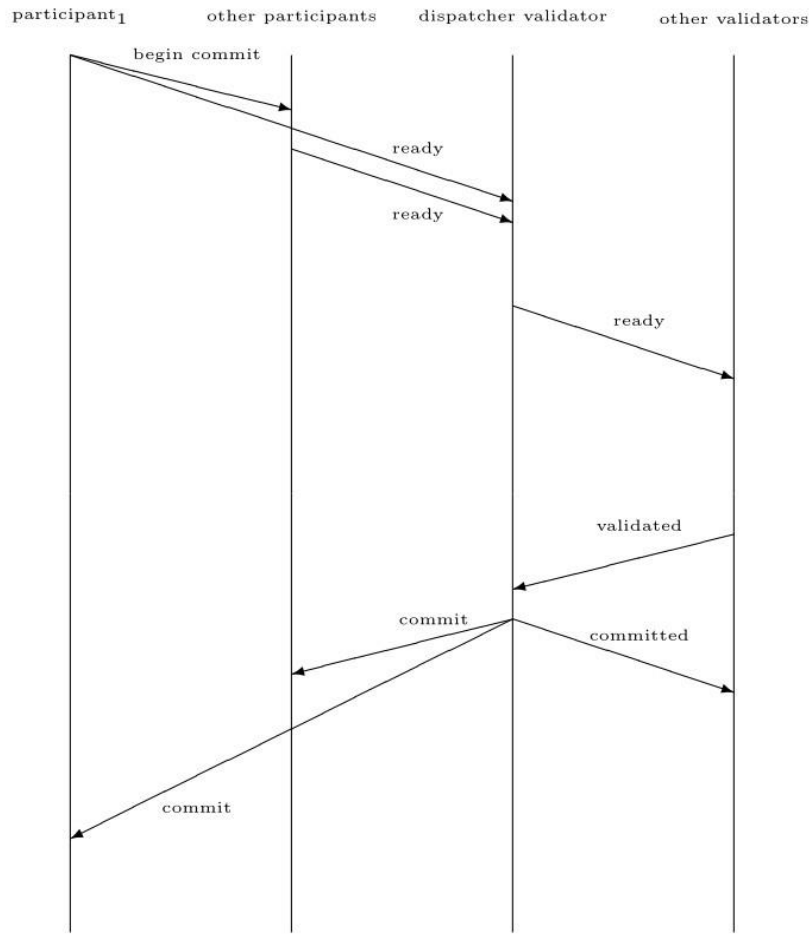

Fig. 1: Consensus messages

## 3. CORRECTNESS

Correctness is an important key concern when talking about consensus algorithms. The formal specification for the proposed algorithm was made using TLA+ language [6].

The model verifies defined invariants in order to test that the algorithm is correct and, as a whole, the specification is intended to serve as the subject of the proof. This also help other people to implement easily and correctly the algorithm in real systems. There may be many causes for failures and maybe some of them can not be tested, but the formal tool help us to analyze all the final states a system can reach in order to identify and resolve potential problems.

### 3.1. CONSENSUS SPECIFICATION

All the actions a node may take are described below:

- **Transaction manager** is the node which initiate the transaction and his special role is to send "Begin" message to all other transaction participants. It is also a participant in transaction, so all the actions below are applicable, except receiving "Begin" message step;
- **Participant nodes**, chronologically are initially in a "working" state. As soon as they receive the "Begin" message from the transaction manager, they move into "preparing" state. During this step, the transaction is locally finalized and every such node ensures that the transaction can be recovered in case of a failure. After all the processings have been done, every participant sends one "Ready" message to dispatcher manager and moves to "ready" state. In this state, each node waits for receiving the commit or rollback decision from the dispatcher node;

- **Dispatcher node** coordinates all validator nodes which work together in order to ensure fault tolerance in case of a dispatcher failure. The node receives "Ready" messages from participants. As soon as such a message is received, it validates locally the message and sends it to other validators. One message is considered validated when validator nodes mark it in majority (in other words, this happens when the dispatcher manager have been received sufficient "Validated" messages). After all the "Ready" messages of a transaction are validated, this node has to send the "Commit" message to all transaction participants. In the end, "Committed" message is sent to other validators in order to mark that the transaction is finished. Of course, the "Rollback" message may be send when not all the messages are validated;
- **Validator nodes** receive from the dispatcher node some "Ready" messages which are first locally validated, then they send "Validated" message back to the dispatcher.

## 3.2. DISPATCHER ELECTION SPECIFICATION

Validating a message means, at least, saving into local memory that message or only the metadata needed for dispatcher failover, which is done using an election algorithm:

- **Coordinator node**: Initially, all validator nodes try to satisfy the launch condition which consist in generating three consecutive numbers greater than a chosen threshold. When this happens, the node sends a "Proposal" message alongside with the greatest random generated number. When the node is voted in majority, it becomes "coordinator" and runs a roulette wheel selection algorithm using the numbers received from other nodes. The winner of this selection will be the leader and its status will be announced to all nodes;
- **Other nodes**: When the "Proposal" message is received, the node votes for sender if it is the first time in the current round of vote and sends his greatest random generated number.

The new dispatcher needs to finalize all pending transactions and this may be a problem unless an additional convention is used. There are some cases which must be analysed:

1) Old dispatcher fails after receiving a certain "Ready" message and sending at least one validation message for that "Ready" message. One of the validators which received the validation message will be elected as the new dispatcher. But the problem is that it does not know anything about other "Ready" messages which might have been sent by other participants and not sent for validation by the old dispatcher. One simple solution is that every participant must send all pending "Ready" messages to the new dispatcher when its announcement is made;

2) Old dispatcher fails before sending to validation the first "Ready" message of a transaction or before receiving the first "Ready" message of a transaction. Of course, the new chosen dispatcher will not know anything about that transaction, so the previous solution could also help in this case.

The conclusion is that there is necessary to add an additional step which consist in sending all pending "Ready" messages from participants to the new dispatcher, when its announcement is made. In this way those transactions can be committed. Initially, in [6], was mentioned an eligibility constraint as only the validator nodes which received the last message sent by the old dispatcher can be valid candidates for leader position; so, an important aspect which appears in this context is that the constraint might be dropped.

## 4. ALGORITHM SAFETY PROOFS

**Definition 1.** Each node's current vote round monotonically increases.

This is straightforward from specification. □

**Definition 2.** There is at most one coordinator in dispatcher election step.

Let's consider there may be two coordinators for the same election round, $C_1$ and $C_2$. This case can appear, of course, when a split vote is happening.

$C_1$ and $C_2$ received the majority of votes, then let $M_1$ be the set of nodes which gave their votes for $C_1$ and $M_2$ the set with all the nodes which voted for $C_2$.

Let node $V$ be $V = M_1 \cap M_2$; this means that V voted for both $C_1$ and $C_2$ ⇒ based on specification, this is *impossible* because V votes only for the first time in a round of vote, so $C_1 = C_2$. □

**Definition 3.** In the end of dispatcher election, only one new dispatcher is chosen.

This results directly from the previous proof as one coordinator will choose only one node as dispatcher, from specification. □

**Definition 4.** The algorithm chooses a dispatcher even $\lceil N / 2 - 1 \rceil$ nodes crash, where $N$ is the total number of validator nodes.

This results from specifications because the leader is chosen by coordinator node, which is elected with the majority of votes from other nodes. If $\lceil N / 2 - 1 \rceil$ nodes crash, there is no problem as majority can still be reached. □

**Definition 5.** The algorithm commits a transaction even $\lceil N / 2 - 1 \rceil$ validator nodes crash.

This is similar with the previous proof as from specifications the transaction is committed when all the "Ready" messages from participants are validated. One message is validated when validator nodes approve it in majority, so the algorithm works fine even $\lceil N / 2 - 1 \rceil$ validator nodes crash because majority can still be reached. □

**Definition 6.** The algorithm commits transactions even the dispatcher fails while processing.

After the current dispatcher fails, a new one is elected and its first task will consist in interpreting the messages it will receive from participant nodes and the pending transactions will continue the commit consensus as previously mentioned. Based on the received list of "Ready" messages, the new leader of validator nodes will know the status of each transaction in order to take all the necessary decisions (for example, send messages to validation or mark a transaction as committed). □

# 5. PERFORMANCE

Performance test was made using 5 nodes running on distinct virtual machines and the consensus for a transaction finished in 235 milliseconds in average, with a minimum of 140 milliseconds and a maximum of 313 milliseconds. In 90% of cases, consensus was reached in at most 289 milliseconds.

More than 1000 concurrent transactions were taken into account. The histogram is shown in figure 2.

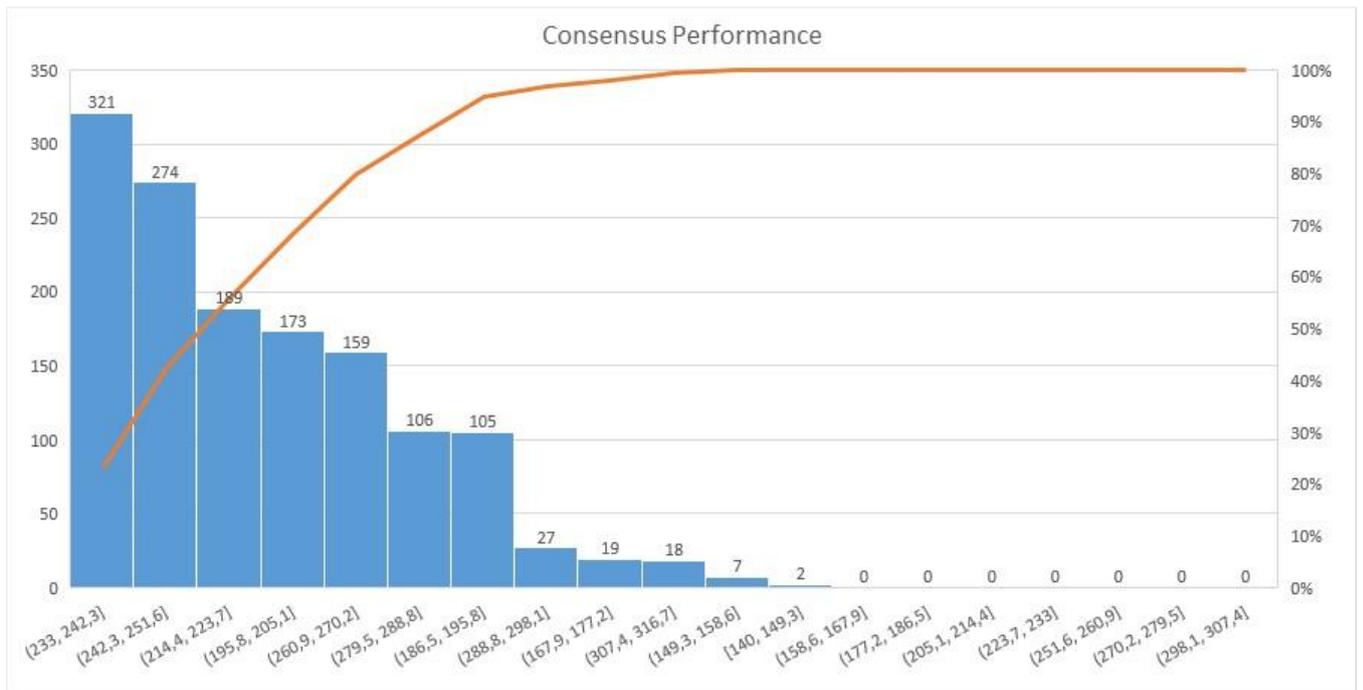

Fig. 2: Consensus performance

# 6. CONCLUSION

The new algorithm analysed in this paper is quite simple and easy to understand. It is correct and safe, proposing a method to solve the consensus agreement problem by using a set of nodes which validate the messages sent between transaction participants and the leader of the validator nodes, called dispatcher validator. It can recover in case this dispatcher node crash and has the capability to continue the pending transactions and commit them eventually.

**Authors**

**Marius Rafailescu** is a Ph.D. candidate at the Department of Computer Science at the "Politehnica" University from Bucharest. His M.S. and B.S were received also from the "Politehnica" University from Bucharest. His main research interests are transactional processing in databases and distributed systems.

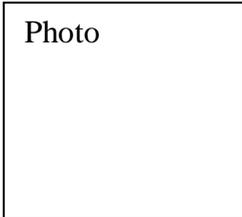
Photo